\newcommand{\be}{\begin{equation}}
\newcommand{\ee}{\end{equation}}
\newcommand{\Tr}{{\rm Tr}}
\def\bea{\begin{eqnarray}}
\def\eea{\end{eqnarray}}
\def\bean{\begin{eqnarray*}}
\def\eean{\end{eqnarray*}}

\newcommand{\barr}{\begin{array}}
\newcommand{\earr}{\end{array}}

\newcommand{\bed}{\begin{displaymath}}
\newcommand{\eed}{\end{displaymath}}
\newcommand{\bal}{\begin{array}{ll}}
\newcommand{\eal}{\end{array}}

\def\bvec#1{\raise1.5ex\hbox{$\rightarrow$}\mkern-16.5mu #1}

\def\thru#1{\mathrel{\mathop{#1\!\!\!/}}}
\def\sthru#1{\mathrel{\mathop{#1\!\!\!\!/}}}
\documentclass{article}
\begin{document}

\centerline{\Large\bf  Dirac's Footsteps and Supersymmetry}
\vskip .2cm

\centerline{P. Ramond}

\vskip .1cm
\centerline
{Institute for Fundamental Theory, Physics Department,} 
\centerline{University of Florida, Gainesville, FL 32611, USA}

\vskip 1cm
 {\obeylines \hfill{\em I am not interested in proofs. 
\hfill I am only interested on how Nature works}
                                                                                                                                  \hfill         P.A.M. Dirac}
\vskip 1cm
\noindent {
One hundred years after its creator's birth, the Dirac equation stands  as the cornerstone of XXth Century physics. But it  is much more, as it carries  the seeds of supersymmetry. Dirac also invented the light-cone, or ``front form" dynamics, which plays a crucial role in string theory and in elucidating the finiteness of $N=4$ Yang-Mills theory. The light-cone structure of  eleven-dimensional supergravity ($N=8$ supergravity in four dimensions) suggests a group-theoretical interpretation of its divergences. We speculate   they  could be  compensated by an infinite number of triplets of massless higher spin fields, each obeying  a  Dirac-like equation associated with the coset $F_4/SO(9)$. The divergences are proportional to the trace over a non-compact structure containing the compact form of $F_4$. Its nature is still unknown, but it could show the way to $M$-theory. }

\section{Dirac's equation}
Much of modern physics starts with the Dirac equation. Its deceptively simple  form  

\be\gamma^\alpha_{}\,p^{}_\alpha\,\Psi~=~\thru{p}\,\Psi~=~0\ ,\ee
obscured the significance of the simple algebraic relation for its anticommutator

\be\{\,\thru{p}\,,\,\thru{p}\,\}~=~p^2_{}\ .\ee
Today we understand this as the seed of supersymmetry, which could be called protosupersymmetry. In the context of string models, this relation turned out to be very potent. There, the momentum was generalized to a momentum density along the string

\be P^\alpha_{}(\sigma)~=~p^\alpha_{}+\mbox {oscillators}\ ,\ee
suggesting that the momentum operator is the average of the momentum density

\be p^\alpha_{}~=~<P^\alpha\,>\ .\ee
 If we assume that the Dirac matrices are themselves like dynamical variables (as envisioned by the Master himself\cite{DIRAC})        

\be \Gamma^\alpha_{}(\sigma)~=~\gamma^\alpha_{}+\mbox{oscillators}\ ,\ee
with 

\be
\gamma^\alpha_{}~=~<\,\Gamma^\alpha_{}\,>\ ,\ee
it is not unnatural\cite{PMR} to expect that 

\be
\gamma^\alpha_{}\,p^{}_\alpha~=~ <\,\Gamma^\alpha_{}\,>\,<P^{}_\alpha\,>~~\rightarrow ~~ <\,\Gamma^\alpha_{}\,P^{}_\alpha\,>\ ,\ee
which generalizes the Dirac equation to the string. If we write

\be
\Gamma^\alpha_{}(\sigma)\,P_\alpha^{}(\sigma)~=~\sum_n\,F^{}_n\,e^{in\sigma}_{}\ ,\ee
we find the superVirasoro algebra

\bea
\{\,F^{}_n\,,\,F^{}_m\,\}&=&2\,L^{}_{n+m}\ ,\cr
[\,L^{}_n\,,\,F^{}_m\,]&=&(2n-m)\,F^{}_{n+m}\ ,\cr
[\,L^{}_n\,,\,L^{}_m\,]&=&(n-m)\,L^{}_{n+m}\ .\eea
We should have read another of Dirac's papers\cite{DIRAC1}on the importance of  $c-$numbers, which were found later by the late Joe Weiss. The rest is history as this algebra opened  the Pandora box of theories with bosons and fermions, which we call supersymmetric.

\section{Divergences and Group Theory}
It is well-known how the running of the couplings in local field theory is related to the divergent part of some diagrams. For example, the one-loop beta function is given by

\be
\beta~=~\frac{11}{3}\,I^{(2)}_{adj}-\frac{2}{3}\,I^{(2)}_f-\frac{1}{3}\,I^{(2)}_H\ ,\ee
where the $I^{(2)}_{adj,\,f,\,H}$ are the quadratic Dynkin indices associated with the adjoint (for the gauge bosons), with spin one-half Weyl fermions, and with complex spin zero fields, respectively. These ``external" group theoretical factors are given by 

\be
\Tr\,(T^A_r\,T^B_r\,)~=~I^{(2)}_r\,\delta^{AB}_{}\ ,\ee 
wher $A,B$ run over the gauge group, and $T^A_r$ are the representation matrices in the $r$ representation. The other numerical factors  stem from group theory of the ``internal" space. As shown by Hughes\cite{HUGHES}, they can be understood for massless particles as, 

\be
\frac{1}{3}\,(1-12 h^2_{})\ ,\ee
where $h$ is the helicity of the particle circulating around the loop. In some sense the square of the helicity can be viewed as the quadratic Dynkin index of the light-cone\cite{DIRAC2} spin group, although in four dimensions, it is only $SO(2)$, which is not much of a group. 

Curtright\cite{CURTRIGHT} generalized this notion when he considered loop integrals coming from theories in higher dimensions, where the spin light-cone little group is more substantial.   
For instance, the one loop vacuum polarization in $N=4$ Yang-Mills can be obtained directly in ten dimensions where the little group is $SO(8)$. Evaluate the loop integrals in four dimensions. The divergent part is proportional to 

\be
(-1)^f_{}\,\Big(\,\frac{1}{3}\,I^{(0)}_{}\,(\delta^{(4)}_{\mu\nu}\,q^2_{}-q^{}_\mu\,q^{}_\nu)-\frac{4}{r}\,I^{(2)}_{}\,(\delta^{(D)}_{\mu\nu}\,q^2_{}-q^{}_\mu\,q^{}_\nu)\Big)\ ,\ee
where  $I^{(0)}$ is the dimension of the {\em transverse little group} representation, $r$ its rank, and $I^{(2)}$ is the quadratic Dynkin index of the same. In the case of $N=4$ Yang Mills, we have $D=10$ and the transverse little group is $SO(8)$. With its triality property, the group theoretical factors are the same for bosons and fermions. This is the genesis of the cancellation of ultraviolet divergences for that theory. It is the peculiar properties of the transverse little group that leads to  ultraviolet finiteness\cite{LARS}, together with supersymmetry in the form of equality of fermions and bosons. This is true for {\it all} higher order Dynkin indices as well: since $SO(8)$ has rank four, it has three more independent Dynkin indices, of order $4,6$, and $8$, which  are the same for bosons and fermions by triality. No other group has that property.

This is puzzling since all string theories stem from eleven-dimensional $M$-theory. The little group there is $SO(9)$, a totally unremarkable group, or so it seems.  In eleven dimensions, supergravity is described by three fields\cite{ELEVEN}, the graviton $h_{\mu\nu}$, a Rarita-Schwinger fermion $\psi_\mu$ and a three-form boson $A_{\mu\nu\rho}$. Their physical degrees of freedom fall in three $SO(9)$ representations whose group-theoretical properties are summarized in the following table:

\hskip 2cm
\begin{center}
\begin{tabular}{|c|c|c|c|}
\hline
$~{\rm Field}~$& $\Psi_i$& $h_{ij}$  & $A_{ijk}$   \\
$~{\rm irrep}~$& $(1001)$&$ (2000)$ & $(0010)$   \\
 \hline 
\hline         
$~I^{(0)}_{}~ $&$  128$ & $ 44$ & $ 84$  \\
 \hline    
$~I^{(2)}_{}~$& $256$& $88$ & $168$  \\
\hline
$~I^{(4)}_{}~$& $640$& $232$ & $408$ \\                                                            
 \hline
$~I^{(6)}_{}~$&$1792$& $712$ &$1080$\\ 
\hline
$~I^{(8)}_{}~$&$5248$& $2440$ &$3000$\\ 
\hline
 \end{tabular}\end{center}
\vskip 0.3cm
 The Dynkin indices have the remarkable property that they cancel between fermions and bosons

\be
I^{(k)}_{\bf 128}~-~I^{(k)}_{\bf 44}~-~I^{(k)}_{\bf 84}~=~0\ ,~~k=0,2,4,6\ ,\ee
except for the highest invariant 

\be
I^{(8)}_{\bf 128}-I^{(8)}_{\bf 44}-I^{(8)}_{\bf 84}~=~-192\ .\ee
This led Curtright to speculate that the theory is divergent because of this inequality. The lowest order divergent diagram that contains the eighth-order invariant  is a three-loop  four-graviton amplitude. While it is hopeless to calculate such a beast, this is the diagram for which there appears to be no local counterterm\cite{DESER}. 

\section{Euler Triplets} 
Sometime ago, it was found\cite{PR} that this pattern of group-theoretical partial cancellations among three representations generalized to other $SO(9)$ representations. 

There are three equivalent embeddings of $SO(9)$ inside the exceptional group $F_4$, much like the I- U- and V-spins for the embedding of $SU(2)\times U(1)$ inside $SU(3)$.  As a result, one can associate with each $F_4$ representation {\it three} $SO(9)$ representations, whose properties are summarized in  the  character formula\cite{GKRS} 

\be
V_{\lambda}\,\otimes\,S^+_{}\,-\,V_{\lambda}\,\otimes\,S^-_{}~=~\sum_{c }\,{\rm sgn}(c)\,U_{c\bullet\lambda}\ .\ee
On the left-hand side,  $V_\lambda$ is a representation of $F_4$ written in terms of its $SO(9)$ subgroup, $S^\pm$ are the two spinor representations of $SO(16)$ written in terms of its anomalously embedded subgroup $SO(9)$, and $\otimes$ denotes the normal Kronecker product of representations. On the right-hand side, the sum is over  $c$, the  three elements of the Weyl group  which map the Weyl chamber of $F_4$ into the (three times larger) chamber of $SO(9)$.  Finally  $U_{c\bullet\lambda}$ denotes the $SO(9)$ representation with highest Dynkin weight $c\bullet\lambda$,  where

$$c\bullet\lambda~=~c\,(\lambda+\rho^{{}^{}}_{F_4})-\rho^{{}^{}}_{SO(9)}\ ,$$ 
and  $\rho$'s are the sum of the fundamental weights for each group, and ${\rm sgn}(c)$ is the index of $c$. Thus to each $F_4$ representation corresponds a triplet, called Euler triplet. The three representations  of supergravity appear in the  trivial case associated with the singlet of $F_4$. Since  

$$SO(16)\,\supset\,SO(9)\ ,\qquad S^+_{}\,\sim\,{\bf 128}={\bf 128}\ ,\qquad S^-_{}\,\sim\,{\bf 128}'={\bf 44}\,+\, {\bf 84}\ ,$$
 the character formula reduces to 

$${\bf 128}\,-\,{\bf 44}\,-\, {\bf 84}~=~{\bf 128}\,-\,{\bf 44}\,-\, {\bf 84}\ .$$
In general, the representations describe (in light-cone variables) fields with spin greater than two. For each  $F_4$ representation with 
Dynkin labels     $[\,a_1\,a_2\,a_3\,a_4\,]$  one obtains  three $SO(9)$ representations listed in order of increasing dimensions:

$$
 (2+a_2+a_3+a_4,a_1,a_2,a_3)\ ,~ (a_2,a_1,1+a_2+a_3,a_4)\ ,~(1+a_2+a_3,a_1,a_2,1+a_3+a_4)$$
For spinor representations, the fourth entry is an odd integer.  Euler triplets for which  the largest  representation  is the spinor have equal   number of fermions and bosons; this occurs whenever both  $a_3$ and $a_4$ are even integers or zero.

\section{Kostant Equation}
We find here again the long hand of Dirac, for the minus sign in the character formula suggests that it is the index formula for a Dirac-like operator. This is Kostant's operator\cite{KOS} associated with the  coset $F^{}_4/SO(9)$. The  Clifford algebra    over this coset

\be
\{\, \Gamma_{}^a\,,\, \Gamma_{}^b\,\}~=~2\,\delta_{}^{ab}\ ,~~a,b=1,2,\dots, 16\ ,\ee
is generated by $(256\times 256)$ matrices, and the Kostant equation is defined as 

\be
{\sthru  K}\,\Psi~=~\sum_{a=1}^{16}\, \Gamma_{}^a\,T^{a}_{}\,\Psi~=~0\ ,\ee
where $T_a$ are the $F_4$ generators not in $SO(9)$, with  commutation relations

\be
[\,T_{}^a\,,\,T_{}^b\,]~=~i\,f^{\,ab\,ij}_{}\,T^{ij}_{}\ .\ee
These are conveniently expressed in terms of copies of 
 $26$ oscillators with the usual Bose-like commutation relations\cite{FULTON}: 
 $A^{[s ]}_0,\; A^{[s ]}_i,\; i=1,\cdots,9,\; B^{[s ]}_a,\; a=1,\cdots,16$, 
and their hermitian conjugates, and where $s =1,2,3 $. 
Under $SO(9)$, the $A^{[s ]}_i$ transform as ${\bf 9}$, $B^{[s ]}_a$ transform as ${\bf 16}$, 
and $A^{[s ]}_0$ is a scalar. Note that the  $B^{[s ]}_a$  satisfy Bose-like commutation relations, even though they are $SO(9)$  spinors. The $F_4$ generators are then   

\begin{eqnarray}
T^{}_{ij}&=&-i\sum_{s =1}^4\left\{\left(A^{[s ]\dag}_iA^{[s ]}_j-A^{[s ]\dag]}_jA^{[s ]}_i\right)+\frac 12\,B^{[s ]\dag}\,\gamma^{}_{ij} B^{[s ]}\label{t_ij}\right\}\nonumber\ ,\\
T_a&=&-\frac{i}{{2}}\sum_{s =1}^4\left\{ (\gamma_i)^{ab}\left(A^{[s ]\dag}_iB^{[s ]}_b-B^{[s ]\dag}_bA^{[s ]}_i\right)-\sqrt{3}\left(B^{[s ]\dag}_aA^{[s ]}_0-A^{[s ]\dag}_0B^{[s ]}_a\right)\right\}\nonumber\ .\label{t_a}
\end{eqnarray}
 One can just as easily have used the coordinate representation of the oscillators by introducing real coordinates $u^{}_i$ which transform as transverse space vectors, $u^{}_0$ as scalars, and $\zeta^{}_a $ as the space spinors. It is amusing to note that the internal cordinates span three exceptional Jordan algebras, which have been the subject of much interest as possible charge spaces.  

The solutions of Kostant's equation are then simply described by a chiral superfield\cite{US}. Listing only its highest weight components, it is of the form 

\be
\Phi_{\vec a}^{}~=~\theta^1\theta^8\,h(y^-,\vec x,\,u_i,\,\zeta_a)~+~\theta^1\theta^4\theta^8\,\psi(y^-,\vec x,\,u_i,\,\zeta_a)~+~\theta^1\theta^4\theta^5\theta^8\,A(y^-,\vec x,\,u_i,\,\zeta_a)\ .\ee
The three fields are polynomials in three sets of ``internal" bosonic coordinates. A possible violation of the spin-statistics connection is avoided when  the fields are even functions of the $\zeta_a$. This happens whenever $a_3$ and $a_4$ are even, but this is the case where each Euler triplet contains as many fermions as bosons. Hence there is an intriguing relation in this solution space between spin-statistics and equality between bosons and fermions. Supergravity is the trivial solution for which the three fields $h\ ,A\ ,$ and $\psi$ are independent of the internal coordinates.

\section{A Zero Sum Game}
 Since the Dynkin indices of the product of two representations satisfy the composition law

$$I_{}^{(n)}[\lambda\otimes \mu]~=~d^{}_\lambda\,I_{}^{(n)}[\mu]+d_\mu^{}\,I_{}^{(n)}[\lambda]\ ,$$
where $d$ is the dimension, it follows that the deficit in $I^{(8)}$ is always proportional to  

$$d_\lambda(I^{(8)}_{S^+}-I^{(8)}_{S^-})\ ,$$
where $d_\lambda$ is the dimension of the $F_4$ representation that generates it. It  is always of the same sign, so cancellation, if it happens will come only after summing over an infinite number of triplets. 

Each triplet contains particles with spin greater than two. As we have shown\cite{US}, they cannot have mass since they do not assemble in massive little group multiplets. The only evasion route from the well documented difficulties~\cite{VASILIEV} with higher spin massless fields interacting with gravity theories is to allow for an infinite number, but to this date no such theory has been put forth, perhaps with good reason. 

If indeed the divergences of supergravity all stem from the deficit in the eighth order Dynkin, we are led to the bizarre equation

\be
\Delta\,I^{(8)}_{}~=~-192\,\sum_{a_1,a_2,a_3,a_4}\,d^{F_4}_{}(a_1,a_2,a_3,a_4)~=~0\ ,\ee
where $d^{F_4}_{}(a_1,a_2,a_3,a_4)$ is the dimension of the $F_4$ representation with Dynkin labels $[\,a_1\,a_2\,a_3\,a_4\,]$.
It can only vanish if the sum is over an infinite number of representations. In order to make sense of this at least two obstacles have to be surmounted. One is to specify the regularization procedure, and the second is to determine the subset of $F_4$ representations over which to take the sum. 

We seek  an algebraic structure which contains an infinite number of $F_4$ representations such that the (regulated) trace, or character, over the dimensions of these representations is zero. It would be wonderful if the triality properties of  affine $E_6^{(1)}$  somehow became important. In  quantum groups, zero  traces can occur when the $q$ parameter is a root of unity\footnote{ I thank E. Mukhin and P. di Francesco for an instructive discussion}, so we are perhaps dealing with such an object. At any rate, it will be a miracle if such a structure exists, but then we are looking for a  unique theory where miracles are expected on a daily basis.  

Dimensions of $F_4$ representations are expressed as $24$th order polynomials in their Dynkin indices. A zero-sum game can be  clearly set up: assuming  $\zeta$-function regularization, or a quantum group, find a subset of $F_4$ representations with zero total dimension.  The existence of such a set would be strong indication for a finite structure underlying eleven-dimensional supergravity.

\section*{Acknowledgments}
I wish to thank Professor H. Baer for his kind hospitality and giving this mere mortal  the honor to speak at the commemoration of one of the Gods of Physics. This work  was supported in part
by the US Department of Energy under grant DE-FG02-97ER41029

\end{document}